\documentstyle[preprint,aps,epsf]{revtex}
\def \k {{\bf k}}
\def \p {{\bf p}}
\def \q {{\bf q}}
\def \z {{\bf z}}
\def \x {{\bf x}}
\def \y {{\bf y}}
\def \hk {{\hat k}}
\def \hp {{\hat p}}
\def \hbk {{\hat{\bf k}}}
\def \hbp {{\hat{\bf p}}}
\begin{document}
\thispagestyle{empty}
\baselineskip = .75\baselineskip


\title{Charge exchange $\rho^0 \pi^+ $ photoproduction 
 and implications for searches of  exotic meson}

\author{Andrei V. Afanasev$^{a,b}$ and Adam P. Szczepaniak$^c$}

\address{
$\makebox{}^a$ Physics Department, North Carolina Central University,
 Durham, NC 27707\\
$\makebox{}^b$ Thomas Jefferson National Accelerator Facility, Newport
 News, VA 23606 \\
 $\makebox{}^c$  Physics Department and Nuclear Theory Center, \\
    Indiana University, Bloomington, IN 47405
}

\maketitle

\begin{abstract}
We analyze the processes $\vec \gamma +p\rightarrow \rho^0 \pi^{+}n$ 
 at low momentum transfer focusing on a possibility of 
  production of an exotic $J^{PC}=1^{-+}$ meson state. 
 In particular we discuss polarization observables
   and conclude that linear 
 photon polarization is  instrumental for separating of the exotic
   wave.

\bigskip 
PACS number(s): 12.39.Mk,
12.40.Nn, 13.60.Le, 13.88+e
\end{abstract}

\narrowtext

\section{Introduction}

 Mesons with unusual quantum numbers 
  play an important role in studies of strong QCD and in 
  understanding of the nature of the effective, low
   energy degrees of freedom. 
   Since, due to their exotic quantum numbers, such 
 mesons cannot be 
   described in terms of the valence
   quarks alone, they in principle give access 
 to the dynamics of the nonvalence
   degrees of freedom thus allowing for studies of strong QCD 
    which go beyond the static $Q\bar Q$ confinement. 
 Recently significant progress has been made in 
  lattice studies of such states and
  several new models have been developed.  
 In these models the  unconventional structure 
  of exotic mesons is typically associated with dynamical gluons
   and lattice gives predictions for 
  the spectrum of gluonic excitations in the presence of static $Q\bar
  Q$ sources. 
 In particular the numerical simulations lead to a series of effective,
     ``excited'' $Q\bar Q$ adiabatic potentials 
  significantly different from that of the  ground state 
 ``Coulomb+linear''. These higher adiabatic potentials 
 arise from gluonic filed configurations  
 which have symmetries distinct from that of the ground state~\cite{lattice}. 
  The adiabatic potentials can then be
   used to predict the spectrum of hybrids with heavy quarks. 

The structure of hybrids containing light quarks 
 is less known. Lattice simulations estimate the mass of the
 ground state   $1^{-+}$ state in the range 
  $1.9 - 2\mbox{ GeV}$~\cite{lattice2} 
  but unlike the  case of 
  heavy hybrids there is not much information available yet about their
  structure~\cite{latt3}.  A number of models have been
 proposed to address this issue. They primarily differ in the 
  treatment of the gluonic degrees of freedom. There are models 
 which describe gluons as 
  quasi-particles {\it i.e.} in a similar way to the constituent 
  quarks~\cite{cons}, other, such as the flux tube model~\cite{flux} 
  associate gluons with 
  collective excitations of a nonrelativistic string as an 
  approximation to the dynamics of the chromoelectric flux between
  the $Q\bar Q$ pair. 
 Finally  
  in the bag model spectrum of the perturbative gluon field
  inside the bag is 
 obtained by imposing boundary conditions on the bag surface~\cite{bag}. 
 All these models lead to modest differences in the predictions for
  the  spectrum and decay pattern of exotic mesons. Future 
  precision data on hybrid production will help 
   discriminating between them and give insight into the
  nonperturbative dynamics of the gluonic field. 

 The experimental studies have not yet resulted in an unambiguous
 spectrum of  exotic mesons~\cite{exp}.  Nevertheless a number of 
 strong candidates have been established. The most recent published  
 analysis of the data from the 
 E852 collaboration at BNL indicates a presence of an exotic
  $J^{PC}=1^{-+}$ signal in the $\rho\pi$ channel of the reaction $\pi^- p \to 
 \rho^0\pi^- p\to \pi^+\pi^-\pi^- p$~\cite{E852}.
 The charged $P$-wave, $(\rho\pi)^{\pm}$ channel
 is particularly useful for exotic searches since it has $G=(-)$ and
 therefore  in the absence of a $I=2$ meson state (which by itself would be
 interesting), belongs to an isovector multiplet
  an therefore has exotic quantum numbers. The $3\pi$ mass spectrum 
(from $\rho^0\pi^{\pm}\to
 \pi^+\pi^-\pi^{\pm}$) is dominated by the $a_1(1270)$, $a_2(1320)$
 and $\pi_2(1670)$ mesons. As discussed in 
  Ref.~\cite{E852} the exotic wave was 
  extracted through partial wave analysis and a 
   state with mass of $1593\pm
 8^{+29}_{-47}\mbox{ MeV}$ and width
 $\Gamma=168\pm20^{+150}_{-12}\mbox{MeV}$ was found to have a resonant
 behavior.

To the best of our knowledge the only photoproduction experiment 
 claiming a possible exotic signal was 
 performed at the SLAC bubble
 chamber with laser backscattered 30 GeV electrons producing  linearly
 polarized photons with an average energy $E_\gamma =19.5\mbox{ GeV}$~\cite{a22}. 
  The measured $3\pi$ mass spectrum looks somewhat different 
  from that produced with the pion beam. Below $M_{3\pi} \sim 1.5 \mbox{ GeV}$ 
 it is still dominated by the
 $a_2$ resonance but there is no clear indication of the $a_1$. At higher mass
  a narrow pick at $M_{3\pi} \sim 1.8 \mbox{ GeV}$ is seen, rather 
 different from the $\pi_2$ 
  seen by the E856 collaboration. A theoretical study of the cross section
 for the photoproduction of a $1^{-+}$ exotic was performed in 
  \cite{ap98} in 
a flux--tube model and finds it to be of the order of 0.5 $\mu
  b$. This is 
 close in magnitude to the cross section of the 
  $a_2(1320)$, and thus in principle, in photoproduction 
 exotic mesons could be produced at a rate similar to the production
 of other, non-exotic states \cite{isgur99}.

As we discuss below the observed photoproduction spectrum is
 consistent with theoretical expectations. In particular, with
 realistic parameters for an $1^{-+}$ exotic state we find that a 
 signal from such a state could stand out above the $\pi_2$ peak  
   in the ${3\pi}$  mass spectrum.  
 Furthermore we discuss the importance of linear photon 
 polarization in the partial wave
 analysis and in particular in isolating the exotic meson signal. 
 This is of central importance for the new  
 generation of meson photoproduction experiments proposed in
 conjunction with the planed energy upgrade at the Jefferson Lab --the
 Hall D project, where production 
mechanisms and decay modes of various meson states could be studied
 with 
 high precision\cite{HallD}. 
The paper is organized as follows. Below we discuss the formalism
 used to describe the photoproduction mechanisms and the treatment of
 the production of the produced $\rho\pi$ pair. In Section.~III 
 we present  our numerical results, and summarize our conclusions 
in Section.~IV.

\section{Formalism}

The process under investigation is 
 $\vec{\gamma} p\rightarrow X^+ n \to (\rho^0 \pi
^{+})n$ and it is shown in Fig.~1.  
 At high photon lab energy, $E_\gamma > \mbox{ few GeV}$ and low 
 momentum transfer, $t=(p'_N -
p_N)^2$, photoproduction is 
  dominated by peripheral production off the meson cloud around the
 proton target.  
 Due to proximity of the pion pole, in a charge exchange reaction
  one expects one-pion-exchange (OPE)  to dominate the production
 amplitude. As discussed in more detail in
 Section~III,  this seems to be supported by the existing data 
 and therefore  
  we will specialize our analysis to this particular case 
  even though the formalism can be 
   easily extended to account for other production mechanisms. 
 One objective of our study will then be to explore the consequences 
 of the unnatural parity exchange (naturality $\tau$ of a particle 
 with spin $J$ and intrinsic parity $\eta$ is defined as 
 $\tau=\eta(-)^J$), 
  for polarization observables. 

The differential cross section for ${\vec\gamma} p\rightarrow X^+ n
\to (\rho^0  \pi
^{+})n$ is given by 

\begin{equation}
{{d^2\sigma }\over {dtdM_{\rho \pi }d\Omega _k }}= {1\over 2}
\sum_{\lambda_N\lambda_{N'}}
{ { 389.3\, \mu\mbox{b GeV}^2}\over {64\pi m_N^2 E_\gamma^2}}
{{|{\bf k}_\rho|}\over {2(2\pi)^3}} 
|A(s,t,M_{\rho\pi},\lambda_\gamma\lambda_N\lambda'_N\lambda_\rho)|^2. 
\end{equation}

\noindent Here $\bf{k}_\rho $ is the 3-momentum of the $\rho^0$ in
the $(\rho^0
\pi^+ )$ c.m.s., $|{\bf k}_\rho| =\lambda(M_{\rho\pi},m_\rho,m_\pi)$ and
  $A$ is the reduced photoproduction amplitude. 
 Processes dominated by a $t$-channel meson exchange are simplest to
 analyze in the Gottfried-Jackson, (GJ) frame defined as the rest frame of the
 produced resonance, $X^+$ (and thus also the $\rho^0\pi^+$ system) 
  with the $z$ axis defined 
 along the direction of the photon momentum and $y$ perpendicular
  to the $\gamma p \to X n$ production plane (see Fig.~2). 
 The spin structure of the photoproduction amplitude 
 $A(s,t,M_{\rho\pi},\lambda_\gamma\lambda_N\lambda'_N\lambda_\rho)$ 
 is in general described 
  in terms of 12 independent complex amplitudes, $i.e.$ 23 real
  functions, however, in the OPE approximation 
 the amplitude $A$ reduces to 
 \begin{equation}
 A =
 A_{OPE}(s,t)\delta_{\lambda_N,-\lambda'_N}T(M_{\rho\pi},t,s,\lambda_\rho\lambda_\gamma). 
 \end{equation}
Here 

\begin{equation}
A_{OPE}(s,t) = \sqrt{2} g_{\pi NN} \left({s\over {s_0}}\right)^{\alpha_\pi(t)}
{{\sqrt{t'} }\over (t - m_\pi^2)}, \label{OPE}
\end{equation}
with $s=2E_\gamma m_N + m_N^2$, 
$\alpha_\pi(t) = 0.9\mbox{ GeV}^{-2}(t-m_\pi^2)$ and 
 the factor of $\sqrt{t'}\equiv \sqrt{|t-t_{\min}|}$ comes from
 the helicity flip $\pi NN$ 
 vertex proportional to the $\pi NN$ coupling given by 
  $g^2_{\pi NN}/4\pi = 14.4$.   
 The amplitude $T$ should be extrapolated off the pion mass shell
 which together with absorption corrections will modify the overall
 $t$-dependence.  We will discuss this in 
 Section~III.  The OPE approximation leaves only
 the amplitude $T(M_{\rho\pi},t,s,\lambda_\rho\lambda_\gamma)$ to be 
 described phenomenologically. Its spin structure is significantly 
 simplified, with only 3 independent helicity amplitudes (or 5 real functions)
 to be determined from measurements. 
 One can demonstrate that
  magnitudes of all of these amplitudes can be extracted 
   from double polarization observables using a linearly polarized
   photon beam and measuring linear polarization of the final $\rho$. 
It
 should be noted that since the spinless particle is exchanged in the 
 $t$--channel,  there is no correlation between the photon and nucleon spins.

\noindent The amplitude $T=T(\gamma \pi^+ \to \rho^0\pi^+)$ will be
constructed below assuming a single resonance dominates a given 
 $J^{PC}$ channel of the $(\rho^0\pi^+)$ system. 
 Phenomenologically, this is expected to be the case for $1\mbox{ GeV}
 < M_{\rho\pi} < 2\mbox{ GeV}$ where the $\rho^0\pi^+$ spectrum 
 seems to be saturated by the $a_1(J^P=1^+)$, $a_2(J^P=2^+)$, and
 $\pi_2(J^P=2^-)$  resonances.  Nevertheless, the formalism 
 developed below can be easily generalized to include other 
 states and possibly a nonresonant  background.

\subsection{Unitary model for the ${\vec \gamma} \pi^+ \to
 \rho^0\pi^+$ amplitude} 
 In order to be able to reproduce full widths of the hadronic resonances 
  contributing to the production of the $\rho^0\pi^+$ pair we have 

  to take into account hadronic states other then just the $\rho\pi$. 
 Fortunately, all of the states listed above 
have typically at most two dominant decay channels and they all 
 correspond to two meson states. 
  For example,  
 almost $100\%$ of the $a_1$ width comes from coupling to the $\rho\pi$
 channel. The $a_2$ decays are also dominated by the $\rho\pi$ mode
($70\%$) followed by the $\eta\pi$ mode ($15\%$). Finally for $\pi_2$
the ratios are, $56\%$ for the $f_2\pi$ decay channel and $31\%$ for the
 $\rho\pi$~\cite{PDG}.  The keep the model simple we will therefore 
  truncate the hadronic Fock space and 
 include a single resonance state $|X\rangle$, and up to two, two-body 
meson channels, $|AB\rangle$, one corresponding to the measured 
 $\rho^0\pi^+$ state and the other specific the the particular
  resonance, as listed above. 
We also introduce a $J^{PC}=1^{-+}$ exotic state and later present 
 numerical results for different 
 choices of its mass and width. 

 In model calculations, an exotic state  with a 
 mass below $2\mbox{ GeV}$ is predicted 
 to couple mainly to the $b_1\pi$ and  $f_2\pi$ followed by the 
 $\rho\pi$ channels. 
   The E852 data,
 however, have so far only seen the exotic wave in 
 the $\rho\pi$ channel~\cite{E852}. It
is possible that final state interactions shift the strength 
 between the different channels and make the couplings to the 
$\rho\pi$ and the $b_1\pi$ channels comparable. 
 We will thus make a crude assumption that  
 decays of the exotic are dominated by a single $S$-wave channel 
 ({\it e.g.} the $(b_1\pi)_S$) 
 and a single $P$-wave in the $(\rho\pi)_P$ channel 
 each contribution roughly $50\%$ to the 
 total hadronic width. 
  The full list of channels used in this study is summarized in Table~1. 

   With the assumptions given above, and ignoring 
   final state interaction (which for example could 
 be incorporated into the scattering matrix in a 
   form of Blatt-Weisskopf  factors)    
    the model is completely specified by a 
   Hamiltonian which in the rest frame of the resonance is given by     
  
\begin{equation}
H = M + \delta M  + K +  V^s + V^e. \label{ham} 
\end{equation}
For the simple channel interactions described above it is also 
 possible to write a manifestly Lorentz invariant expression for the
 resulting $S$-matrix. 
  We choose to work with a fixed frame Hamiltonian formalism to enable
 a simple connection with other noncovariant models, 
 {\it e.g.} quark model, flux tube model or  bag model 
 where most of the calculations for the channel couplings, 
  especially involving hybrids have been made. 
  In Eq.~(\ref{ham}) the first term represents the 
 kinetic energy of a resonance with a physical mass 
 $M|X\rangle=m_X|X\rangle$ and the bare mass $(M + \delta M)|X\rangle=(m_X+\delta m_X)
|X\rangle \equiv {\bar m}_X|X\rangle$. The second term 
 is the  kinetic energy of the 
 two mesons $K|AB,\k\rangle = (\sqrt{m_A^2 + \k^2} + \sqrt{m_B^2 + \k^2})|AB,\k\rangle$ 
 with $\k$ being the relative 3-momentum between the two mesons 
 in their rest frame. The potentials 
 $V^s$ and $V^e$ describe strong and electromagnetic
 couplings between the resonance $X$ and the two-meson states,  respectively. 
  The mass shifts $\delta m_X$ will be adjusted to match positions 
   of the maxima of the scattering amplitude $T$ with the physical
 masses, $m_X$ of the resonances,  and the  
  strong couplings specifying $V^s$ will be fixed by the 
   phenomenological strong partial widths of the resonances. 
 With absence of other channels coupled to the two-meson state 
 there are no mass shifts to $m_A$ or $m_B$. 
  Also, to first order in the electromagnetic interactions the strength of
$V^e$ can be directly calculated from the $X\to \gamma\pi$ decay widths. 

 We will first discuss the electromagnetic interactions. 
The truncated matrix elements of $V^e$ ({\it e.g.} with the total 
3-momentum conserving 
$\delta$-functions eliminated) in the resonance rest frame are given by
\begin{equation}
\langle \gamma\pi;\q,\lambda_\gamma|V^e|X;J^P,\lambda_X\rangle = 
 \sum_{L_\gamma}(m _X g^{L_\gamma}_{X\gamma\pi}(q) ) \left({q\over
 {m_X}}\right)^{L_\gamma} V^e_{J^PL_\gamma}(\hat \q,\lambda_X,\lambda_\gamma), \label{Ve}
\end{equation}
where ${\bf q}$ is the photon momentum in the $\gamma\pi$ rest frame,
 $q=|\q|$, 
 $\lambda_\gamma$ is the photon helicity,  and the resonance spin,
$\lambda_X$ is quantized along the $\z$ axis. 
Furthermore, $V^e_{J_X^PL_\gamma} = \sum_\lambda \sqrt{{2L_\gamma+1}\over {4\pi}}
\langle J_X\lambda_X|L_\gamma 0,1\lambda\rangle
  D^{J_X}_{\lambda\lambda_\gamma}(\hat{\bf q})$ 
 and the Wigner rotation takes care of the possible difference in the 
 direction of photon polarization and spin quantization 
 axis if the photon momentum is not parallel to the later. 
 In the GJ frame, however, 
 $\q = q\hat \z$ and the Wigner rotation becomes trivial, 
 $D^{J_X}_{\lambda\lambda_X}(\hat\z) = \delta_{\lambda\lambda_X}$.  
 The momentum dependence of the electromagnetic 
  couplings,  $g^{L_\gamma}_{X\gamma\pi}(q)$ could be 
  calculated in  a given microscopic model. 
 
The matrix elements of the strong interaction potential are given by 
 
\begin{equation}
\langle X;J^P,\lambda_X| V^s | AB; \k,
s_A\lambda_A,s_B\lambda_B\rangle =  
\sum_{L_{AB}} 
({\bar g}^{L_{AB}}_{XAB}(k) m_X) 
\left( {k\over {m_x}}\right)^{L_{AB}} V^s_{J^PL_{AB}}(\hat{\k},
\lambda_Xs_As_B\lambda_A\lambda_B),  
\end{equation}
with $V^s_{J^PL_{AB}} = \sum_{M_{AB}s_{AB}\lambda_{AB}}
 \langle s_{AB}\lambda_{AB}|s_A\lambda_A,s_B\lambda_B\rangle
\langle
s_{AB}\lambda_{AB},L_{AB}M_{AB}|J\lambda_X\rangle 
Y_{L_{AB}M_{AB}}({\hat\k})$. 
We choose a simple parameterization of the 
 momentum dependence of the bare couplings 
  ${\bar g}^L_{XAB}(k)  = {\bar g}^L_{XAB}\exp(-k^2/\Lambda^2)$ 
  with $\Lambda = O(1\mbox{ GeV})$.  
It will be convenient to define 
$q(M)\equiv\lambda(M,0,m_\pi)$ and $k_{AB}(M)\equiv\lambda(M,m_A,m_B)$
 which represent the breakup momenta  of the $\gamma\pi$ and of the $AB$
  system respectively coming from a decay of a state with mass $M$.  

The $\gamma\pi \to AB$ scattering amplitude, $T$ is obtained from 
  the solution of the Lippmann-Schwinger equation for the Hamiltonian
  in  Eq.~(\ref{ham}) and it is given by,  

\begin{equation}
\langle \gamma \pi; \q,\lambda_\gamma|T(E)|
AB;  s_A\lambda_A,s_B\lambda_B\rangle 
= \sum_{X;J^P\lambda_X L_\gamma L_{AB}}  V^e_{J^PL_\gamma}(\q,\lambda_X,\lambda_\gamma)
T_{X,J^P}^{L_\gamma L_{AB}}(E)
V^s_{J^PL_{AB}}(\k,\lambda_Xs_As_B\lambda_A\lambda_B),  
\end{equation}
with 
\begin{equation}
T_{X,J^PL_\gamma L_{AB}}(E) =  
 \left({q\over {m_X}}\right)^{L_\gamma}
 { {  m_X g^{L_\gamma}_{X\gamma\pi}(q) {\bar g}^{L_{AB}}_{XAB}(k) }
\over {  2 \left( E - m_X  -
  \Sigma_{X,J^P}(E) + i {{ \Gamma_{X,J^P}(E)}\over 2}\right)} }
  \left({k\over {m_X}}\right)^{L_{AB}}, \label{Tampl}
\end{equation}
where the energy dependent shifts in the real part are given by  
\begin{eqnarray}
\Sigma_{X,J^P}(E) & = & \sum_{AB,L_{AB}}
{{({\bar g}^{L_{AB}}_{XAB}(k_{AB}(m_X)) m_X)^2}\over {64\pi^3}} \nonumber \\
& \times & {\cal P}\int_{m_A+m_B} {{dM}\over {M}}\left({k_{AB}(M)\over m_X}\right)^{2L_{AB}+1}
  \left(  { { {\bar g}^{L_{AB}}_{XAB}(k_{AB}(E)) }\over 
{ {\bar g}^{L_{AB}}_{XAB}(k_{AB}(m_X)) }}\right)^2 
\left({1\over {E-M}} - {1\over {m_X-M}}\right).
\end{eqnarray}
The second term under the integral 
 comes form the $\delta m_X$ mass counter-term chosen so that 
 $\Sigma(m_X)=0$ making the 
  real part of $T$ vanish at the position of the resonance. 
   The  
 energy dependent widths are 
\begin{eqnarray}
\Gamma_{X,J^P}(E) &  = &  \sum_{AB,L_{AB}} \Gamma^{L_{AB}}_{X,J^P}(E) = 
\sum_{AB,L_{AB}}
m_X {{({\bar g}^{L_{AB}}_{XAB}(k_{AB}(E)))^2}\over {32\pi^2}}{{m_X}\over {E}}
\left({k_{AB}(E)\over {m_X}}\right)^{2L_{AB}+1} \nonumber \\
&  = &
 \sum_{AB,L_{AB}} \Gamma^{BW, L_{AB}}_{X,J^P}\left( { {k_{AB}(E)}\over {k_{AB}(m_X)}}\right)^{2L_{AB}+1}
 {{m_X}\over E} \left(  { { {\bar g}^{L_{AB}}_{XAB}(k_{AB}(E)) }\over 
{  g^{L_{AB}}_{XAB}(k_{AB}(m_X))  }}\right)^2, 
\end{eqnarray}
 where $\Gamma^{BW, L_{AB}}_{X,J^P}$ is the Breit-Wigner (BW) width of the resonance $X$ 
  decaying into a two-body hadronic state $AB$ in the $L_{AB}$ partial wave. 
   The physical,  
 $g^{L_{AB}}_{XAB}$ and bare couplings, ${\bar g}^{L_{AB}}_{XAB}$ are 
related by 
 
 \begin{equation}
 g^{L_{AB}}_{XAB} = {{ {\bar g}^{L_{AB}}_{XAB} } \over 
  { \left[1 - \left({d\over
  {dE}}\Sigma_{X,J^P}(E)\right)_{E=m_X}\right]^{1/2} } },  
 \label{coup} \end{equation}
 so that near the resonance 
 pole $T_X(E)$ has the BW form,  

\begin{equation}
T_{X,J^PL_\gamma L_{AB}}(E\sim m_X) =  
 \left({{q(m_X)}\over {m_X}}\right)^{L_\gamma}
 { {  m_X g^{L_\gamma}_{X\gamma\pi}(q_X) g^{L_{AB}}_{XAB}(k_X) }
\over {  2 \left( E - m_X  
  + i {{ \Gamma^{BW}_{X,J^P}}\over 2}\right)} }
  \left({{k_{AB}(m_X)}\over {m_X}}\right)^{L_{AB}}. 
\end{equation}
Table~2 lists the numerical values of the bare and physical couplings
obtained for the set of resonance 
parameters given in  Table~1.  
 For the $AB=\rho^0\pi^+$ final state, the unpolarized 
 differential cross section integrated 
 over the solid angle $\Omega_k$ in the GJ frame is then given by 
 
  \begin{equation}
{{d^2\sigma }\over {dtdM_{\rho \pi } }}= 
{ { 389.3\, \mu\mbox{b GeV}^2}\over {4m_N^2 E_\gamma^2}}
|A_{OPE}(s,t)|^2
 {M_{\rho\pi}\over q}
\sum_{X,J^P} {{ m_X\Gamma_{X\to\gamma\pi}(q(M_{\rho\pi}))
\Gamma^{L_{\rho\pi}}_{X\to\rho^0\pi^+}(k_{\rho\pi}(M_{\rho\pi})) }
  \over {(M_{\rho\pi} - m_X - \Sigma_{X,J^P}(M_{\rho\pi}))^2 
   + \left({{\Gamma_{X,J^P}(M_{\rho\pi})}\over 2}\right)^2 }
  }. 
\end{equation}
 The energy dependent electromagnetic widths are calculated from,
 \begin{eqnarray}
& & \Gamma_{X\to \gamma\pi}(E) =  m_X \sum_{L_\gamma,L'_\gamma}
{{g^{L_\gamma}_{X\gamma\pi}(q(E))
    g^{L'_\gamma}_{X\gamma\pi}(q(E))}\over {32\pi^2}}
 \left({{q(E)}\over {m_x}}\right)^{L_\gamma+L'_\gamma+1} \nonumber \\
& \times & \left[\delta_{L_\gamma,L'_\gamma}\left(\delta_{L_\gamma,J} + \delta_{L_\gamma,J+1}{J\over {2J+1}}
+ \delta_{L_\gamma,J-1}{{J+1}\over {2J+1}}\right) + 
\left(\delta_{L_\gamma,J+1}\delta_{L'_\gamma,J-1} +
   \delta_{L'_\gamma,J+1}\delta_{L_\gamma,J-1}\right)
{{\sqrt{J(J+1)}}\over {2J+1}}\right]. \nonumber \\ \label{photo}
\end{eqnarray}
To account for the kinematics of the off-shell pion in the $t$-channel
  we replace the momentum 
$q(M_{\rho\pi})$ in the the angular momentum factors, by 
$q(M_{\rho\pi}) = \lambda(M_{\rho\pi},0,m_\pi) \to
\lambda(M_{\rho\pi},0,t)$. We however, keep the on-shell 
$q(M_{\rho\pi})$ in the argument of the 
 couplings, $g^{L_\gamma}_{X\gamma\pi}(q(E))$.  
 We will further study the resulting $t$-dependence of the full 
 photoproduction amplitude in Section~III. 

   Unlike the case of strong decays considered here, 
  in radiative decays (to real photons), 
    $L_\gamma$ is not a good quantum number and as seen from Eq.~(\ref{photo}), 
     if more then one partial amplitude contributes  
 the radiative width, $\Gamma_{X\to\gamma\pi}$ does not determine 
  the individual electromagnetic couplings
  $g^{L_\gamma}_{X\gamma\pi}$ 
   but only a 
  linear combination of products. 
  In the case considered here  $a_1$ and $\pi_2$ have more then
  one  partial wave contributing 
  to the $\gamma\pi$ decay, the $S+D$ and $P+F$ waves respectively. 
 We will fix the ratio of these amplitudes to match the ratio of the
   corresponding waves in the $X\to \rho\pi$ decays as in the 
  VMD model.

\subsection{Polarization observables} 

It is well known that 
linear polarization gives access
 the the largest possible number of independent production amplitudes. Even more importantly, however, 
 linear polarization is necessary for isolating $t$-channel 
 natural from unnatural parity exchanges \cite{schilling}.  
 This follows from transformation properties of the production
 amplitude 
 under parity which relates  amplitudes with opposite
 helicities and depends on the naturalities of particles involved in
 the process. 
 It then follows that in order to be able to discriminated between different
  naturalities it is necessary to have a {\it coherent} superposition
  of helicity states, which in the case of real photons corresponds to
  linear, or more generally elliptical polarization.  
 Alternatively, if the naturality of the $t$-channel exchange is
 presumed to be 
  known, as in the case studied here, parity arguments 
 imply that the correlation between polarization direction of 
  the photon and that of the produced mesonic resonance will depend on
  the naturality of the produced 
  resonance. The proof is straightforward. Consider the matrix element of the 
 electromagnetic interaction $V^e$ 
   given in Eq.~(\ref{Ve}). 
 Parity invariance  restricts the possible
  values of $L_\gamma$ so that $(-1)^{L_\gamma}
= \eta_X$, where $\eta_X$ is the intrinsic parity of the
resonance. From the properties of the CG coefficients it follows that such a
  matrix element satisfies,

\begin{equation}
V^e_{J^PL_\gamma}(\lambda_\gamma,\lambda_X) = (-1)^{J_X + L_\gamma+1}
V^e_{J^PL_\gamma}(-\lambda_\gamma,-\lambda_X)
 =  -\tau_X V^e_{J^PL_\gamma}(-\lambda_\gamma,-\lambda_X), \label{par}
\end{equation}
 where $\tau_X= \eta_X (-1)^{J_X}$ is the naturality of the
 resonance.   

A photon beam linearly polarized either along $\y$ 
 {\it i.e.} perpendicular to 
 the production plane or along $\x = \y \times \z$ corresponds to an 
 initial state $| i\rangle$, 
  $ i=\x,\y$ which is 
  a linear superposition of helicity states $|\lambda_\gamma\rangle$, 
 $\lambda_\gamma = \pm 1$, 

\begin{equation}
|{\bf x}\rangle = {1\over {\sqrt{2}}}\left( |-1\rangle - |+1\rangle\right)\,,\;\; 
|{\bf y}\rangle = {i\over {\sqrt{2}}}\left( |-1\rangle + |+1\rangle\right)\,\;\; ,
\end{equation}
Introduction a similar basis to describe the orientation of the linear 
polarization of 
 the resonance and rewriting the matrix elements of $V^e_{J^PL_\gamma}$
 given by Eq.~(\ref{par}) in this basis leads to 

\begin{equation}
V^e_{J^PL_\gamma}({\bf x}_\gamma,{\bf x}_X) = V^e_{J^PL_\gamma}({\bf y}_\gamma,{\bf y}_X) \ne 0,\;\;
V^e_{J^PL_\gamma}({\bf y}_\gamma,{\bf x}_X) = V^e_{J^PL_\gamma}({\bf x}_\gamma,{\bf y}_X) = 0,\;\; \label{polu}
\end{equation}
if the produced resonance is unnatural ($\tau_x = -1$) and 
\begin{equation}
V^e_{J^PL_\gamma}({\bf y}_\gamma,{\bf x}_X) = V^e_{J^PL_\gamma}({\bf x}_\gamma,{\bf y}_X) \ne 0,\;\;
V^e_{J^PL_\gamma}({\bf x}_\gamma,{\bf x}_X) = V^e_{J^PL_\gamma}({\bf y}_\gamma,{\bf y}_X) = 0,\;\; \label{poln}
\end{equation}
if it is  natural ($\tau_X = +1$), respectively. 
In other words, if OPE  dominates production 
  the direction of linear 
  polarization of the produced resonance will be  parallel to that of the 
  incoming photon if the produced resonance is unnatural and
 perpendicular if it is natural. 
 As will be shown below this provides an important tool for isolating 
 the produced resonances. 

Polarization observables may simplify due to  simple spin structure of OPE. 
In particular, one may show from parity arguments that the photons polarized perpendicular
 to the production plane (i.e., along $y$-axis), would not lead to linear polarization
of the final $\rho$ along $z$-direction. This would not be true if a vector particle is exchanged
in $t$--channel.
 
 The measurement of the direction of linear polarization 
 of the resonance requires analysis of the angular distribution of its
 decay products. Recall that in the case of a vector resonance
 decaying  into two (pseudo)scalars 
 the so called $\Sigma$ asymmetry is introduced to measure the degree
 of linear polarization of the vector meson. 
 It is defined as 
 
 \begin{equation}
 \Sigma =  {  {W_{\y}({\pi\over 2},{\pi\over 2}) - 
  W_{\x}({\pi\over 2},{\pi\over 2}) } \over 
 { W_{\y}({\pi\over 2},{\pi\over 2}) + 
 W_{\x}({\pi\over 2},{\pi\over 2})  } }
\end{equation}
where the intensities $W_i(\theta_k,\pi_k)$ come from the photons
polarized 
  in the direction $i=\x,\y$ and  
 $\theta_k$ and $\phi_k$ are the decay angles representing direction of flight 
 of one of the two (pseudo)scalars, $\k = k(\sin\theta_k\cos\phi_k,\sin\theta_k\sin\phi_k
  ,\cos\theta_k)$  in the GJ frame. 
 The decay amplitude of a spin-$1$ resonance 
 linearly polarized along $i_X = \x_X,\y_X$ 
  to two (pseudo)scalars if proportional to 
  $\sum_{\lambda_X=\pm 1}\epsilon^{i_X}(\lambda_X) 
 Y_{1,\lambda_X}(\theta_k,\phi_k) $ 
 where $\bbox{ \epsilon}(\lambda_X)$ is the resonance
 polarization vector. This results in an amplitude proportional to 
 $k^x$ or $k^y$ for vector mesons  polarized along $\x$ and $\y$ respectively. 
 In other words, vector meson $100\%$ polarized along 
   the $\x$ or $\y$ direction leads to an intensity of the decay products 
     picking along $\x$ or  $\y$ direction respectively. 
  If the correlation between photon and resonance polarization 
 comes from Eq.~(\ref{poln})   
  ({\it i.e.} a natural resonance is produced 
  via an  unnatural $t$-channel exchange {\it e.g.} OPE) 
   then  polarization of the vector meson is 
  perpendicular to that of the photon. Since $\Sigma$ refers to decay distributions measured 
  along $\y$ ($\theta_k=\phi_k=\pi/2$) a nonvanishing contributions
  comes from photons polarized along $\x$ resulting in $\Sigma=-1$. 
 Alternatively, if production is via a natural exchange 
 ({\it e.g.} Pomeron) then 
  vector meson polarization is parallel to that  
  of the photon and non-vanishing $\Sigma$ comes from photons
  polarized  along $\y$ 
  resulting in $\Sigma=+1$. 
 In the general case when both natural 
 and unnatural 
  $t$-channel exchanges take place simultaneously, in vector meson production 
   $\Sigma$ asymmetry can be used to 
 discriminate between the two mechanisms.

 In the case under study, the photoproduced resonance 
 decays into a vector ($\rho^0$) and a 
 pseudoscalar ($\pi^+$)  and 
 it is necessary to generalized the definition of the asymmetry 
  in order to reflect the correlation between photon and resonance 
 polarizations. 
 The $\rho^0\pi^+$ decay distribution depends on the polarization of the 
 $\rho^0$, which in turn is 
 reflected in the angular distribution  of the $\pi^+\pi^-$ from its 
  decay. Thus it is best to study a distribution which
  correlates the 
  direction of the relative momentum, 
   $\k$ between the $\rho^0$ and the $\pi^+$ coming from the decay of
  the photoproduced resonance 
     and the direction the relative momentum, $\p$ (measured in the 
 rest frame of the $\rho^0$)   between 
  the two pions from the $\rho^0$ decay. 
  To obtain the amplitude $T$ describing the 
  $\pi^+\pi^-\pi^+$ production we multiply the amplitude in
 Eq.~(\ref{Tampl})  
   (for $AB=\rho^0\pi^+$) 
  by $\p{\bbox \epsilon(\lambda_{\rho^0})}/\sqrt{2}$
   corresponding to a BW approximation to the $\rho^0$ 
  propagator (the factor of $\sqrt{2}$ coming from the isospin). 
 Summing over $\lambda_{\rho^0}$ allows to express 
 the  $3\pi$ photoproduction in terms of the four angles
  $\theta_k,\phi_k$, and $\theta_p,\phi_p$ specifying the
   directions of ${\hat\k}$ and ${\hat\p}$ respectively. 
  In Table~3 we give the analytical forms of these dependencies 
 for the 
   resonances considered here. 
   Inspecting the formulas in Table~3 it follows that in particular 
  measuring $d\sigma/dM_{3\pi}d\Omega_p d\Omega_k$ 
 at $\theta_p = \pi/4$, $\phi_p=\pi/2$ and $\theta_k = \pi/4$,
  $\phi_k=3\pi/4$  eliminates 
 contribution from the $\pi_2$ resonance. Thus if the exotic state is 
  weakly produced near the $\pi_2$ mass region linearly polarized
   photons enable to enhance the signal in the partial wave
  analysis.  

\section{Numerical Results}
 We first discuss the momentum transfer dependence of $d\sigma/dt$. 
The existing data on $\rho^0\pi^+$ photoproduction in the mass  
region of interests, ($1 \mbox{ GeV}<M_{3\rho^0\pi^+} < 2\mbox{ GeV}$),
comes from two, SLAC 
experiments~\cite{a22,a21}. The measurement of Eisenberg {\it et al.} 
 was performed at two photon energies, $E_\gamma=4.3$ and  $5.25 \mbox{ GeV}$. 
 Reconstructed $3\pi$ mass spectrum shows a clear signature of the $a_2$ 
 resonance and a broad enhancement in the mass region of $M_{3\pi}=1.5-2\mbox{ GeV}$. 
 The total $a_2$ photoproduction cross section, (averaged over the two 
 photon energies) was found to be $0.9 \pm 0.6 \mu b$. This, however, as 
  pointed out in Ref.~\cite{a22}, 
 does not account for other decay modes of the $a_2$ correcting for
 which  gives  
     $\sigma(\gamma p \to a_2^+ n)  = 2.6 \pm 0.6 \mu b$ 
   at $E_\gamma = 4.8\mbox{ GeV}$. 
 In Ref.~\cite{a21} the momentum transfer dependence has been fitted to  
  OPE with absorption corrections corresponding to final state  $a_2 n$ 
  scattering. These corrections have been calculated using the 
  strong cutoff model (SCM) where partial waves of the 
   OPE amplitude are cut from below and contributions from waves with 
  $J<J_c$ are eliminated. 
  The cutoff parameter, $J_c$ was fitted to data and found to
   correspond  to an 
  absorption radius of $\sim 1\mbox{ fm}$.  
  The second experiment of Condo 
  {\it at al.}  measured the $\rho^0\pi^+$ photoproduction at the
   average 
 photon energy of 
  $E_\gamma = 19.5 \mbox{ GeV}$ and for the total $a^+_2$
   photoproduction
 cross section 
   finds $\sigma(\gamma p \to a_2^+ n) = 0.29 \pm 0.09 \mu b$   
 At these higher photon energies $t$-dependence was also found to be 
 consistent with OPE, and it was parameterized with 
  a single exponential form, $d\sigma/dt 
 \propto \exp(b t)$
  with $b\sim 10\mbox{ GeV}^{-2}$. 
  
 In Fig.~3 we show the results of the different parameterizations of the 
 $t$-dependence in the  $a^+_2$ mass region 
  at $E_\gamma=4.8\mbox{ GeV}$ as compared to the data from Ref.~\cite{a21}, 
 and   rescaled as described above. 
  To account for the absorption corrections we have replaced the 
 OPE amplitude of   Eq.~(\ref{OPE}) by

\begin{equation}
A_{OPE}(s,t) \to A_{OPE}(s,t)T_{cor}(t), \label{abs} 
\end{equation}
  with the correction term parameterized by 
 \begin{equation}   
T_{cor}(t) = (a_1 e^{b_1 t} + a_2 e^{b_2 t})^{1/2}. \label{tcor}
\end{equation}
To obtain the $a^+_2$ photoproduction cross section we have integrated the 
differential cross section $d\sigma/dtdM_{\rho^0\pi^+}$ calculated
 using the formalism described in Sec.II 
 over the $a_2$ mass region corresponding to the data of Ref.~\cite{a21}, 
{\it i.e}  $M_{a^+_2} - \Gamma/2 < M_{\rho^0\pi^+} < M_{a^+_2} + 
\Gamma/2$ with  $M_{a_2} = 1.31 \mbox{ GeV}$ and $\Gamma = 80\mbox{
  MeV}$. The parameters entering in Eq.~(\ref{tcor}) have been chosen to
 give the best description of the $E_\gamma = 4.8\mbox{ GeV}$ and the
 $E_\gamma=19.5\mbox{ GeV}$ data of  Condo {\it et al.}, leading to 
  $\sigma(\gamma p \to a_2^+ ) 
  = 1.92 \mu b$ and  $\sigma(\gamma p \to a_2^+ n) 
  = 0.43\mu b$ for the two energies respectively. This corresponds 
  to  $b_1 \sim 35\mbox{GeV}^{-1}$, 
  $b_2 \sim 3\mbox{ GeV}^{-2}$, $a_1\sim 30.0$ and $a_2 \sim 1.5$ for
  $\Lambda=1\mbox{ GeV}$.
  The $t$ dependence at $E_\gamma=4.8\mbox{ GeV}$ is shown by the solid line 
  in Fig.~3. The calculation at the higher energy 
 takes  into account the energy dependence of the OPE amplitude as
 well as the different $a^+_2$ mass region quoted in
 Ref.~\cite{a22}, 
     $M_{a_2} = 1.325 \mbox{ GeV}$ and 
  $\Gamma = 150\mbox{ MeV}$. 
  We have also studied dependence of the form factor scale $\Lambda$. 
  For smaller cutoff we find that it becomes 
 harder to simultaneously reproduce 
  the cross sections at both energies
   $4.8$ and $19.5\mbox{ GeV}$. Specifically,  for $\Lambda=0.5\mbox{ GeV}$ the 
   predictions are $\sigma = 1.86 \mu b$ and $\sigma = 0.46 \mu b$ 
 respectively. 
 On the other hand for cutoff larger then $1\mbox{GeV}$ our simple model 
  with one 
  resonance in a given partial wave becomes inadequate as indicated by a  
   rapid increase of the mass shifts ${\bar m}_X - m_X \sim 300-500 
 \mbox{MeV}$.    
  In Fig.~3 we also show $d\sigma/dt$ 
   calculated with the pure OPE amplitude ({\it i.e} using $T_{cor} =
   1$) (dashed line). Finally the straight dotted line corresponds to 
    the parameterization of Condo {\it et al} {\it i.e.} with a
   replacement 
 
    \begin{equation}
    A_{OPE} \to Ae^{bt/2}
    \end{equation}
   with $b\sim 10\mbox{ GeV}^{-2}$ and normalized to match to total 
 cross section 
    calculated with $T_{cor}$ of Eq.~(\ref{tcor}).  

 From the results  in Fig.~3 it is clear that the single exponential 
  does not reproduce the entire $t$-dependence as good as the 
  two-exponential parameterization of Eq.~(\ref{tcor}) indicating that the absorption corrections are 
  indeed significant.   
 We should stress, however, that the form given in Eq.~(\ref{tcor}) is only 
 a parameterization and does not correspond to a particular absorption model. 
 In fact as seen from Fig.~3 
  at low-$t$ the 
  data is above the OPE prediction by a factor of $2-3$ {\it i.e} $T_{cor}>1$ 
  while for a truly absorptive correction 
  one should have $T_{cor}<1$. Using a specific absorption model 
{\it e.g.} SCM,   $T_{cor}<1$ is obtained automatically 
  but then  to reproduce the magnitude of the cross section 
  requires couplings ({\it e.g.} 
    $g_{a_2\gamma\pi}$) significantly larger than
  quoted by the PDG~\cite{PDG}. This is in fact what happens with the fit in 
    Ref.~\cite{a21} where the extracted $\Gamma_{a_2\to \gamma\pi}$ is 
     larger then used here by almost a factor of two. 
    Furthermore the form of OPE cross section  used in Ref.~\cite{a21} 
    is enhanced at low $t$ since it was chosen proportional 
  to $t$ rather then $t'$, which all together makes the 
     the fit to their data possible with an absorption term, $T_{cor}<1$. 
 More precise data is clearly needed to resolve these discrepancies. 

  In the following we will continue with the 
   parameterization of Eq.~(\ref{tcor}) with the set of parameters listed
  above. In Fig.~4 we show the mass dependence of
 $d\sigma/dM_{\rho^0\pi^+}$  when only the single $a^+_2$ 
 resonance is retained in the photoproduction amplitude. 
 The upper 
solid line uses $\Lambda=1\mbox{ GeV}$ and the dashed line which is
 lower  at larger $M_{\rho^0\pi^+}$) is the 
is the corresponding BW approximation. The second 
 solid line and the corresponding BW approximation 
 (dashed line) correspond to $\Lambda=0.5\mbox{ GeV}$. In 
 Fig.~5 we give the 
 full prediction of the model {\it i.e.} with all resonances
 included.  In Fig.~5a the $J^{PC}=1^{-+}$ resonance mass 
 has been set at $M=1.775\mbox{ GeV}$ which corresponds to the Condo
 state, 
and the different plots correspond to 
$\Gamma=100\mbox{ MeV}$ and $\Gamma^e=400\mbox{ keV}$ (upper 
 solid line), $\Gamma=100\mbox{ MeV}$ and $\Gamma^e=200\mbox{ keV}$ (lower 
 solid line), $\Gamma=200\mbox{ MeV}$ and $\Gamma^e=400\mbox{ keV}$ (upper 
 dashed line), $\Gamma=200\mbox{ MeV}$ and $\Gamma^e=200\mbox{ keV}$ (lower
 solid line). In Fig.~5b we set $M_{1^{-+}}=1.6\mbox{ GeV}$ as
 measured by the 
 E852 collaboration, with $\Gamma=170\mbox{ MeV}$ and 
 $\Gamma^e=400\mbox{ keV}$ (solid line), $\Gamma^e=600\mbox{ keV}$ (dotted line), 
 and $\Gamma^e=200\mbox{ keV}$ (dashed line). 
 Finally in  Fig.~6 we show the sensitivity to the degree of photon 
 linear polarization defined as

 \begin{equation}
 \delta {{d\sigma}\over {dM_{\rho^0\pi^+}d\Omega_{k}d\Omega_{p}}} 
  = {1\over P_\gamma}\left({{d\sigma}_{\y}\over {dM_{\rho^0\pi^+}d\Omega_{k}d\Omega_{p}}} 
   - {{d\sigma}_{\x}\over {dM_{\rho^0\pi^+}d\Omega_{k}d\Omega_{p}}} \right)
 \end{equation}
 where the subscript in $\sigma$ refers to the direction of the 
 photon polarization and 
 $0\le P_\gamma \le 1$ is the degree of linear polarization. 
 In Fig.~6a we choose $|k_x|=|k_y|$, $\hbp\cdot\hbk=0$ and  
 $\theta_p = 0.35\pi$. It follows from the angular distributions listed in
  Table~3, that at these angles contributions to the asymmetry from 
  $|\pi_2|^2$ and $|a_1|^2$ identically vanish 
  leaving only the the $a_2$ and $1^{-+}$ intensities together with 
 all  interference terms. 
  The solid line is the full calculation with all resonances included
  while  the dashed line 
   has the $1^{-+}$ exotic state removed. In Fig.~6b 
  choosing $\theta_p= \pi/4$, $\phi_p = \pi/2$, 
 $\theta_k=\pi/4$, $\phi_k=3\pi/4$ 
   which leads to $\hp_y\hk_z + \hp_z\hk_y=0$ ,$\hk_x=\hp_x=0$ and 
   $\hbp\cdot\hbk=0$ we 
 have  removed the entire contribution from the $\pi_2$
 (including all interferences with this wave). 
   The solid line corresponds the full calculation (solid line) and the
 pick at $M_{3\pi}\sim 1.6\mbox{ GeV}$ comes almost entirely from the
 exotic wave since $\pi_2$ contribution has been removed. 
 When the exotic is not put into the calculation (dashed line) the remaining
 small  contribution comes from the broad background from the $a_1$. 
 
   \section{Conclusions}

 In this paper we studied the reaction $\vec\gamma p \to X^+ n
  \to \rho^0\pi^+ n \to \pi^+\pi^-\pi^+ n$ for linearly polarized
   photons at low momentum transfer. We find a qualitative
   agreement between the data and a theoretical 
 description based on the one pion exchange 
 mechanism, however, more quantitative analysis reveals presence of
  corrections from absorption and possibly other production mechanisms.  
 The expected
  dominance of the OPE which fixes naturality in the $t$-channel
  enables, through polarization observables to discriminate between
  naturalities of the produces resonances. Even without full partial
  wave analysis it turns out to be possible to find maxima in the
  angular distribution of the $\pi^+\pi^-\pi^+$ system which are dominated by a
   single resonance. In particular, for exotic, $1^-+$ meson production we
  have shown that there are regions 
  where the contribution from the $\pi_2$ meson can be completely
  eliminated leaving the possibility for the $1^{-+}$ to peak in the
  intensity. There are other directions where, for example, intensities of the $\pi_2$ and $a_2$
  mesons do not contribute. From the formulas in Table~3 more such regions
    can be found. 
  Of course, this type of analysis cannot be a substitute
  for the full partial wave analysis, in particular since, as
  discussed above, other production mechanisms, {\it e.g.} a natural $\rho$
  exchange can modify the angular distributions. 
  Nevertheless, to the extent OPE does dominate the low-$t$, charge exchange
  photoproduction the analysis given here provides a 
  simple filter of the exotic wave.

\section*{Acknowledgments}

We thank the members of Hall D Collaboration for stimulating
discussions. This work supported by the DOE under contracts, 
DE--AC05--84ER40150 and 
     DE--FG02--87ER40365.  



\begin{table}
\begin{tabular}{lcccccc}
Resonance & $J^{PC}$   & Mass[GeV] & Partial waves & $\Gamma^{L_{AB}}_{XAB}/
\Gamma$ &  $\Gamma$[MeV]  & $\Gamma^e$[keV] \\
decay channel & & & & & &   \\ \hline 
$a_1$                           &  $1^{++}$  & 1.26        &   & & 400 & 640    \\
$\;\;\;\;\;\;\rho\pi$           &            &             & S & 0.99      & & \\
                                &            &             & D & 0.01      & &\\ \hline 
$a_2$                           &  $2^{++}$  & 1.32        &   & & 110 &  295   \\
$\;\;\;\;\;\;\rho\pi$           &            &             & D & 0.70      & & \\
$\;\;\;\;\;\;$rest ($\eta\pi$)  &            &             & D & 0.30      & & \\ \hline 
$\pi_2$                         & $2^{-+}$   & 1.67        &   & & 258 & 300    \\
$\;\;\;\;\;\;\rho\pi$           &            &             & P & 0.98*0.31 & & \\
                                &            &             & F & 0.02*0.31 & & \\
$\;\;\;\;\;\;$rest ($f_2\pi$)   &            &             & S & 0.69      & & \\ \hline
$\hat\rho$                      & $1^{-+}$   & 1.6-1.8     &   & & 100-200 & 200-600 \\
$\;\;\;\;\;\;\rho\pi$           &            &             & P & 0.50      & &  \\
$\;\;\;\;\;\;$rest ($b_1\pi$)   &            &             & S & 0.50      & &  \\ \hline
\end{tabular}
\caption{Resonance parameters used in the model. The numerical values for the   
 widths are taken from the PDG. $\Gamma$ is the total hadronic width, 
  $\Gamma = \sum_{AB,L_{AB}}\Gamma^{L_{AB}}_{XAB}$ 
   and $\Gamma^e$ is the total radiative width to $\gamma\pi$.
}

\label{table I}
\end{table}

\begin{table}
\begin{tabular}{lcccc}
Resonance & Partial wave & ${{(g^{L_{AB}}_{XAB}(k_{AB}(m_X)))^2}\over {4\pi}}$ & 
${{({\bar g}^{L_{AB}}_{XAB}(k_{AB}(m_X)))^2}\over {4\pi}}$ &  
 ${{(g^{L_\gamma}_{X\gamma\pi})^2}\over {4\pi}}$   \\
decay channel & & & &   \\ \hline 
$a_1$                           &   &        &        &        \\
$\;\;\;\;\;\;\rho\pi$           & S & 26.28  & 18.70  &  $1.03$      \\
                                & D & 32.54  & 23.15  &  $1.04\times 10^{-2}$      \\ \hline 
$a_2$                           &   &        &        &        \\
$\;\;\;\;\;\;\rho\pi$           & D & 443.4  & 563.48 &  $2.37$      \\
$\;\;\;\;\;\;$rest ($\eta\pi$)  & D & 57.08  &  72.71 &        \\ \hline 
$\pi_2$                         &   &        &        &        \\
$\;\;\;\;\;\;\rho\pi$           & P & 20.01  & 15.54  & $1.74$   \\
                                & F & 20.15  & 15.64  & $4.01\times 10^{-2}$ \\
$\;\;\;\;\;\;$rest ($f_2\pi$)   & S & 13.72  & 10.65  &        \\ \hline
$\hat\rho$                      &   &        &        &        \\
$\;\;\;\;\;\;\rho\pi$           & P & 24.59  & 20.85  & $1.51$   \\
$\;\;\;\;\;\;$rest ($b_1\pi$)   & S &  7.12  & 6.04   &        \\ \hline
\end{tabular}
\caption{Strong, physical and physical and electromagnetic couplings as defined in text. 
 The strong couplings are evaluated for $\Lambda=1\mbox{ GeV}$.  
 The physical couplings are obtained from the BW widths given in Table~1 and then 
  the bare couplings are calculated using Eq.~(\ref{coup}). For the $1^{-+}$ we used 
   $\Gamma = 170\mbox{MeV}$ and $\Gamma^e = 400\mbox{keV}$. } 
\label{table II}
\end{table}

\begin{table}
\begin{tabular}{lcc}
Resonance & Partial wave & $W(\theta_k,\phi_k,\theta_p,\phi_k)$ \\ \hline
$a_1$                           & S  &  $\hp^2_y$               \\
                                & D  & $\left(\hk_y\hbp\cdot\hbk - {{\hp_y}\over 3}\right)^2$                \\ \hline
$a_2$                           & D  
& $\left(\hk_z [\hbk\times\hbp]_x + \hk_x [\hbk\times\hbp]_z\right)^2$  \\ \hline
$\pi_2$                         & P  & $(\hp_y\hk_z + \hp_z\hk_y)^2$                \\
                                & F  & $\left(\hk_z\hk_y(\hbp\cdot\hbk)-{{\hk_z\hp_y+\hp_z\hk_y}
                                \over 5}\right)^2$              \\ \hline
$\hat\rho$                      & P  &  $\left([\hbk\times\hbp]_x\right)^2$              \\ \hline
\end{tabular}
\caption{Unnormalized angular decay distributions 
of the $\pi^+(-\p)\pi^-(\p)\pi^+(\k)$  for photons polarized in the
direction perpendicular to the production plane ({\y}). The angular
distributions corresponding to photons polarized along $\x$ are
obtained by interchanging the subscripts $y\to x$ and $x\to y$ in the
third column. }
\label{table III}
\end{table}


\hbox to \hsize{%
\begin{minipage}[t]{\hsize}
\begin{figure}
\epsfxsize=3in
\hbox to \hsize{\hss\epsffile{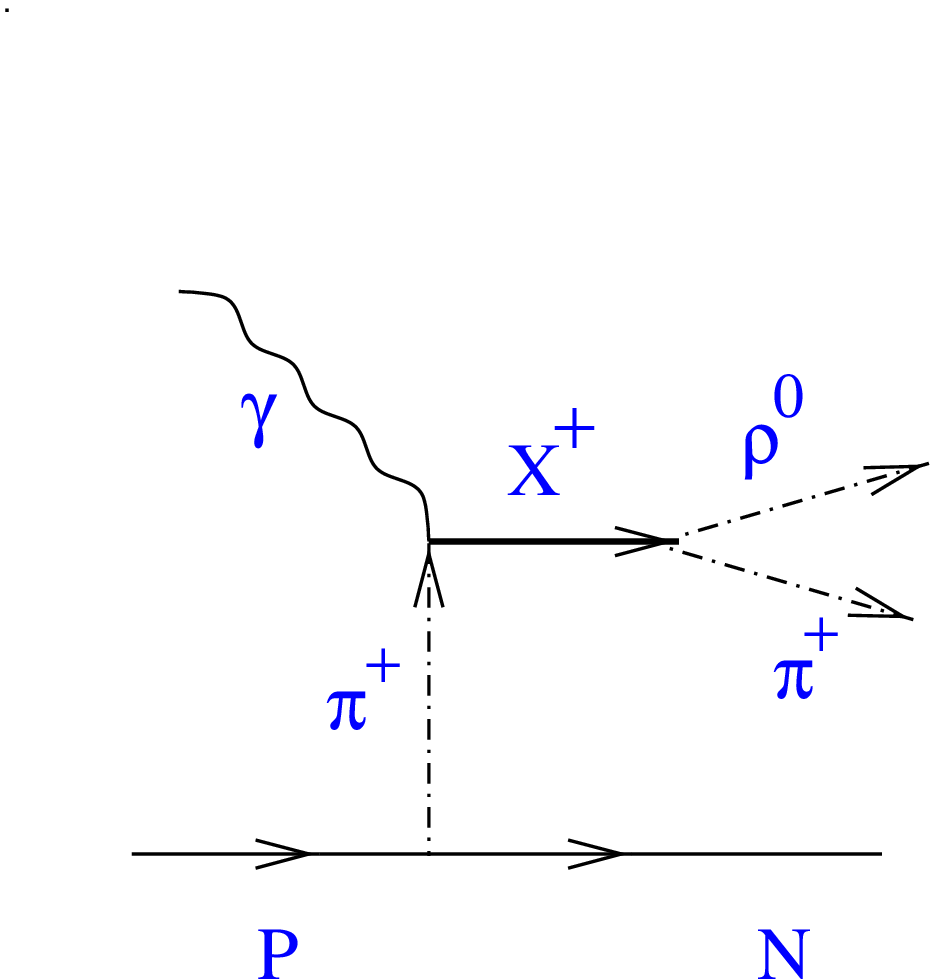}\hss}
\label{fig:amb}
\end{figure}
\end{minipage}}
\begin{center}
  {\small Fig.~1.  $\rho^0\pi^+$ photoproduction via one-pion-exchange. 
   }
\end{center}

\hbox to \hsize{%
\begin{minipage}[t]{\hsize}
\begin{figure}
\epsfxsize=5in
\hbox to \hsize{\hss\epsffile{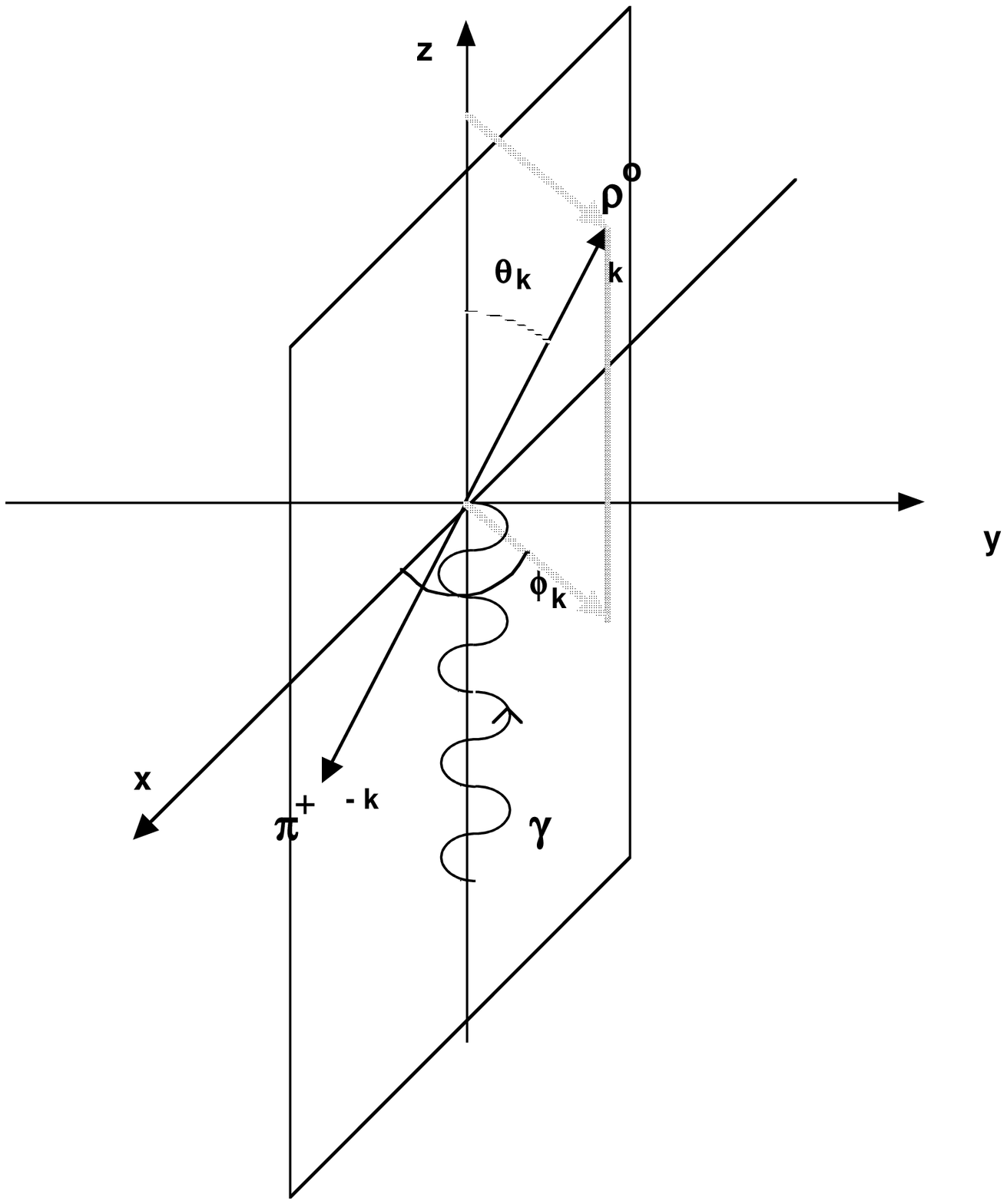}\hss}
\label{fig:gj}
\end{figure}
\end{minipage}}
\begin{center}
  {\small Fig.~2. Kinematics of the $\rho^0\pi^+$ production in the GJ frame. }
\end{center}

\hbox to \hsize{%
\begin{minipage}[t]{\hsize}
\begin{figure}
\epsfxsize=5in
\hbox to \hsize{\hss\epsffile{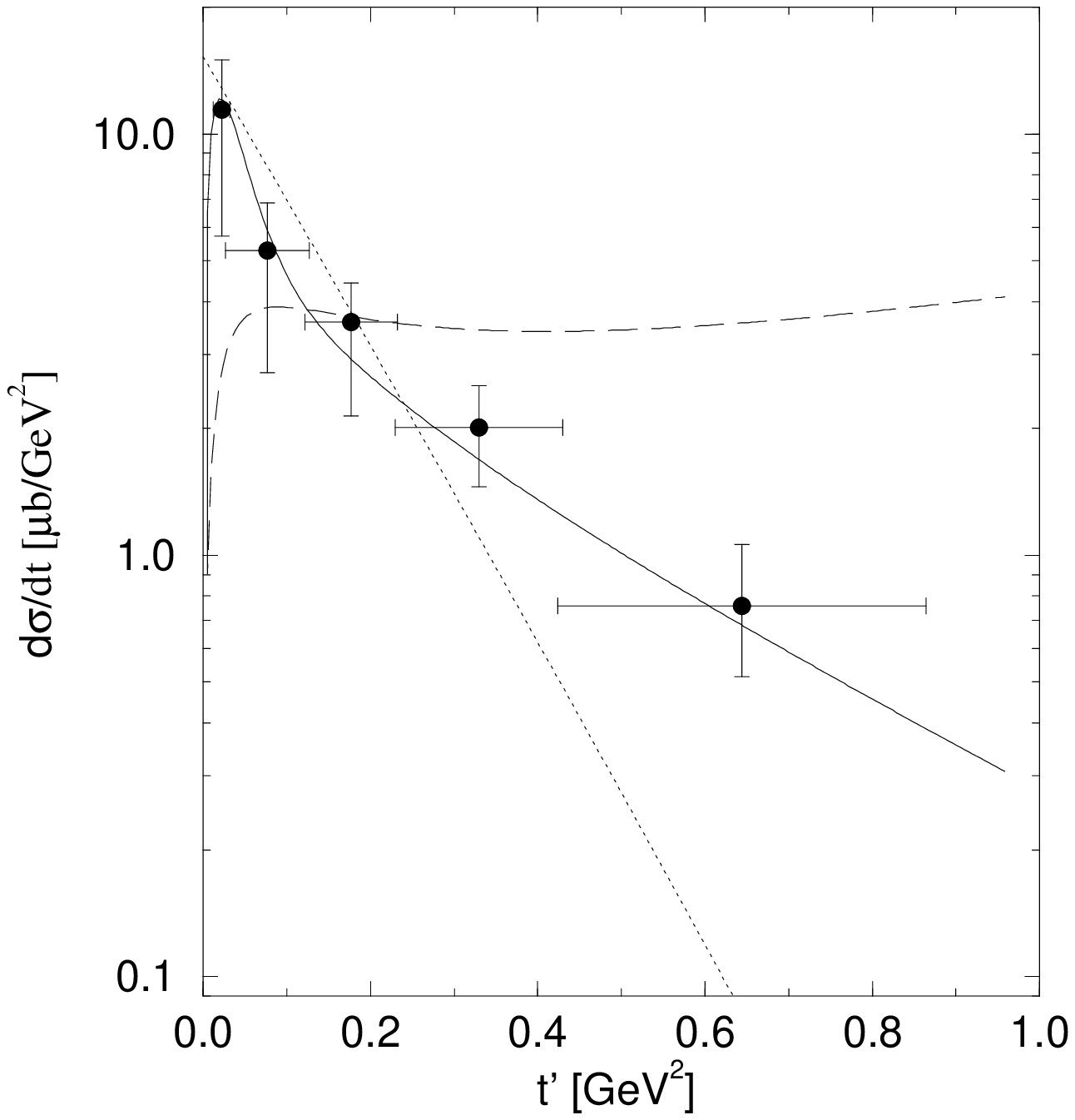}\hss}
\label{fig:dsdt}
\end{figure}
\end{minipage}}
\begin{center}
  {\small Fig.~3. Momentum transfer dependence of $a_2^+$ photoproduction cross section. 
  Data is from Ref.~\cite{a21}. Solid line is the OPE prediction 
  corrected to account for absorption.   
   Dashed line is the pure OPE 
     prediction and the dotted line is the $A\exp(-b t)$ parameterization. }
\end{center}

\hbox to \hsize{%
\begin{minipage}[t]{\hsize}
\begin{figure}
\epsfxsize=5in
\hbox to \hsize{\hss\epsffile{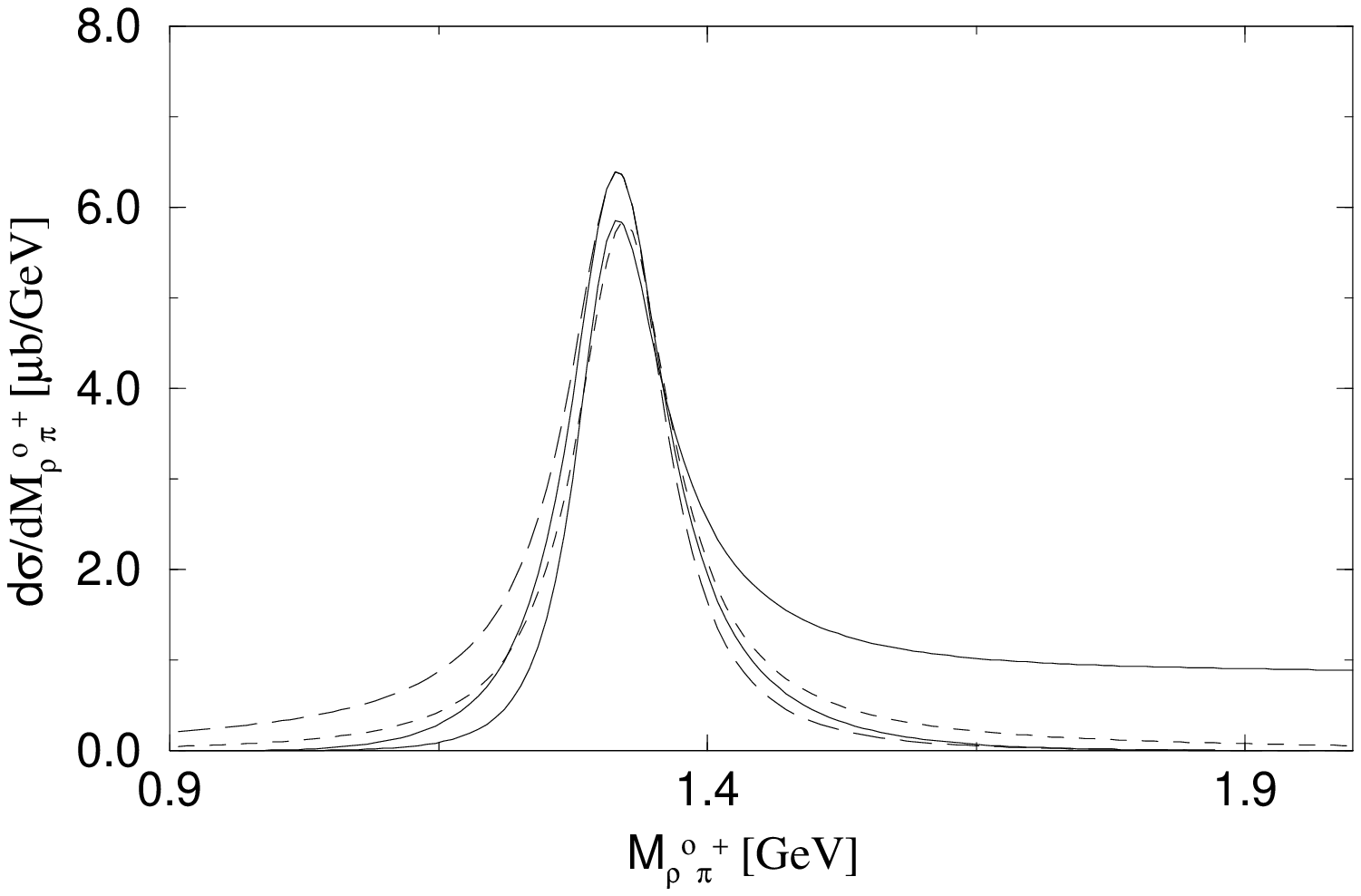}\hss}
\label{fig:a2}
\end{figure}
\end{minipage}}
\begin{center}
  {\small Fig.~4. $\rho^0\pi^+$ photoproduction cross 
  section through a single $a^+_2$ resonance. 
   The various curves  are explained in the text. } 
\end{center}

\hbox to \hsize{%
\begin{minipage}[t]{\hsize}
\begin{figure}
\epsfxsize=5in
\hbox to \hsize{\hss\epsffile{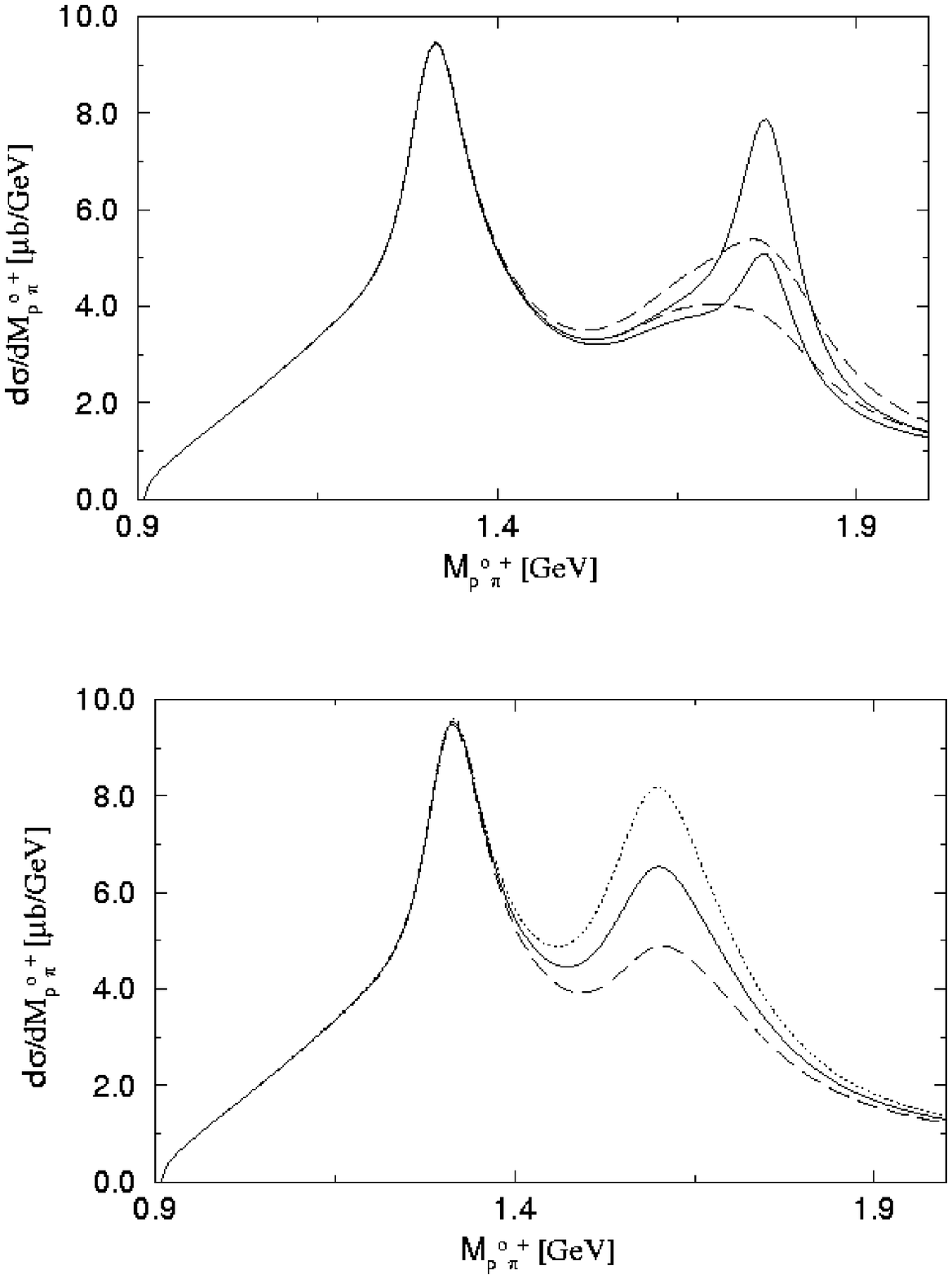}\hss}
\label{fig:all}
\end{figure}
\end{minipage}}
\begin{center}
  {\small Fig.~5a,b. $\rho^0\pi^+$ photoproduction cross 
  section via with the contributions from $a_1$, $a_2$, $\pi_2$ and a 
   $1^{-+}$ exotic state for different widths of the 
    exotic resonance for a) $M_{1^{-+}}=1.775\mbox{ GeV}$ and b) 
      $M_{1^{-+}}=1.6\mbox{ GeV}$. }
\end{center}

\hbox to \hsize{%
\begin{minipage}[t]{\hsize}
\begin{figure}
\epsfxsize=5in
\hbox to \hsize{\hss\epsffile{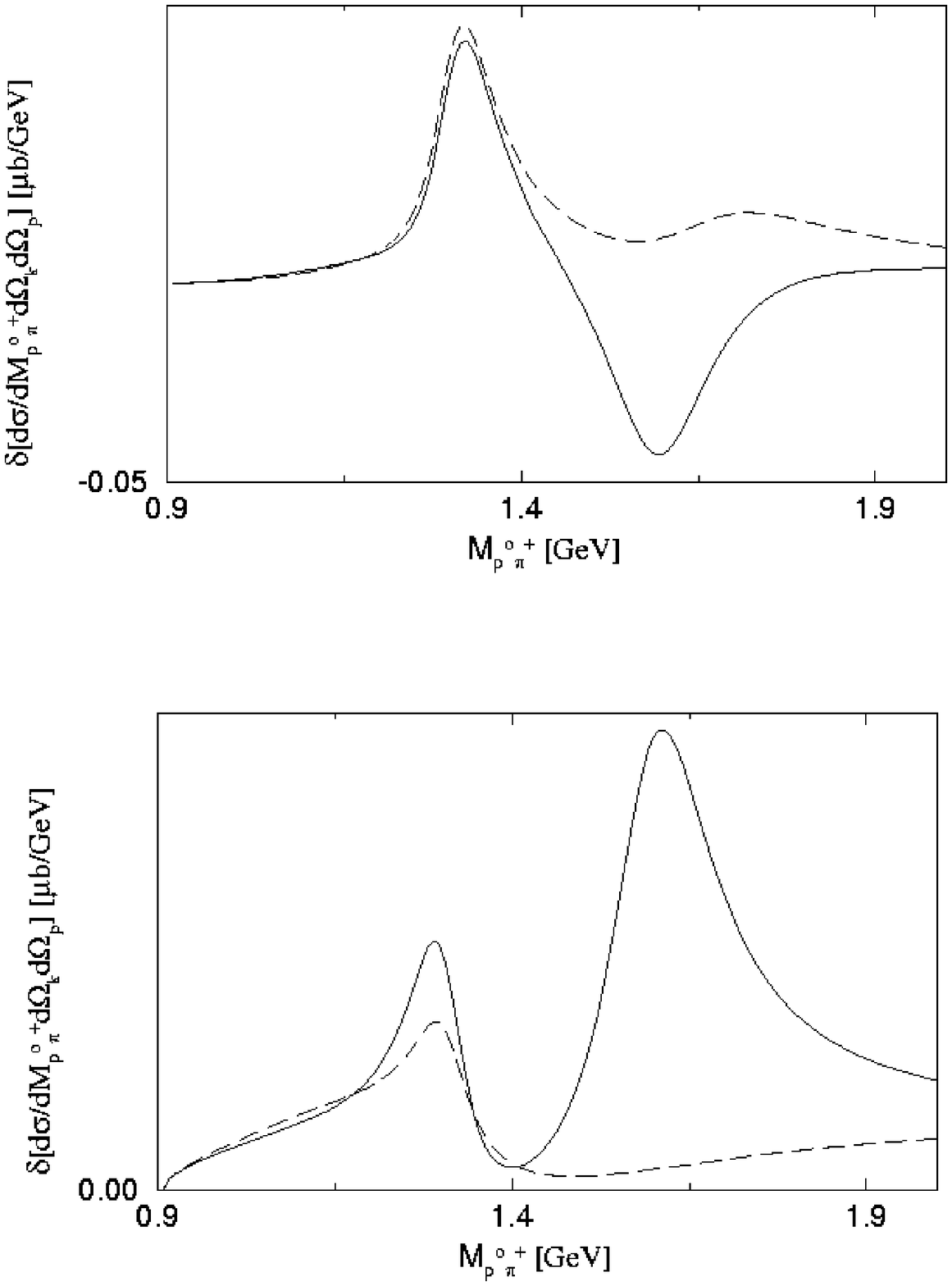}\hss}
\label{fig:pol}
\end{figure}
\end{minipage}}
\begin{center}
  {\small Fig.~6a,b. $\delta d\sigma/dM_{\rho^0\pi^+d\Omega_pd\Omega_k}$ 
   for linearly polarized photons along-\x and along-\y at a)
    $|k_x|=|k_y|$, $\hbp\cdot\hbk = 0$, $\theta_p=0.35\pi$ and b) 
    $\hbp\perp\hbk$, $k_x=0$, $\theta_k=\pi/4$. } 

\end{center}


\begin{references}

\bibitem{lattice}  K.J.~Juge, J.~Kuti, C.J.~Morningstar, 
    Nucl. Phys. Proc. Suppl. {\bf 63},  326 (1998); C.J.~Morningstar, 
K.J.~Juge, J.~Kuti, LANL e-print archive hep-lat/9809098. 

\bibitem{lattice2} P.~Lacock, K.~Schilling, LANL e-print archive
  hep-lat/9809022; C.~Bernard {\it et al.} Phys. Rev. D{\bf 56}, 7039
  (1997); C.~Bernard {\it et al.} LANL e-print archive
  hep-lat/9809087; P.Lacock, C.~Michael, P.~Boyle, P.~Rowland,
  Phys. Lett, B{\bf 401}, 308 (1997).


\bibitem{latt3} M.~Hess, F.~Karsch, E.~Laermann, I.~Wetzorke,
  Phys. Rev. D{\bf 58}, 111502 (1998). 

\bibitem{cons}  D. Horn, J. Mandula, Phys. Rev. D{\bf 17}, 898 (1978);
   K.~Yamada, S.~Ishida, H.~Takahashi, {\it in proceedings of }
     HADRON 95, The 6th International Conference on Hadron
   Spectroscopy,  Manchester, Jul 10-14, 1995; 
  A.P.~Szczepaniak, E.S.~Swanson Phys.~Rev. D{\bf 55}, 3987 (1997). 


\bibitem{flux} N.~Isgur, R.~Kokoski, J.~Paton, Phys. Rev. Lett. {\bf
    54}, 869 (1985); N.~Isgur, J.~Paton Phys. Rev. D{\bf 31}, 2910
    (1985);  E.S. Swanson, A.P.~Szczepaniak. Phys. Rev. D{\bf 59}, 
 014035 (1999); 
 P.R.~Page, E.S.~Swanson, A.P.~Szczepaniak, Phys. Rev. D{\bf 59},
   034016 (1999). 

\bibitem{bag} T.~Barnes, F.E.~Close, F.de~Viron, J.~Weyers, 
  Nucl. Phys. B{\bf 224}, 241 (1983); P.~Hasenfratz, R.R.~Horgan,
  J.~Kuti, J.M.~Richard, Phys. Lett. B{\bf 95}, 299 (1980); 
 S.~Ono, Z. Phys. C{\bf 26}, 307 (1984). 


\bibitem{exp} A.~Palano,  Nucl. Phys. Proc. Suppl. BC {\bf 39}, 287,
  (1995); M. Kunze {\it et al.} Phys. Atom. Nucl. {\bf 57}, 1497 (1994)
 

\bibitem{E852} G.S.~Adams, {\it et al.} Phys. Rev. Lett. {\bf 81}, 5760.
  (1998). 
  
\bibitem{a22}  G.T.~Condo {\it et al.}, Phys. Rev. {\bf D}48, 3045 (1993).

\bibitem{ap98} A. Afanasev, P.R. Page, Phys. Rev. {\bf D57}, 6771 (1998).

\bibitem{isgur99} N.Isgur, Preprint JLab-THY-99-09, 16pp; E-print Archive: hep-ph/9904494.

\bibitem{HallD} A.Dzierba {\it et al.} The Hall D Design Report, http://dustbunny.physics.indiana.edu/HallD. 

\bibitem{PDG} C.~Caso {\it et al.}, Eur. Phys. J.  C{\bf }3, 1
                             (1998). 

\bibitem{schilling} K.~Schilling {\it et al.}, Nucl.Phys. {\bf B15}, 397 (1970).


\bibitem{a21} Y.~Eisenberg {\it et al.}, Phys. Rev. Lett. {\bf 23}, 1322 (1969).





\end{references}
\end{document}